\documentclass[letter]{aa} 

\usepackage{graphicx}
\usepackage{txfonts}
\usepackage{amsfonts}
\begin{document} 
	\title{The tidal heating of the exoplanet 55 Cnc e. The role of the orbital eccentricity.}
	\author{Sylvio Ferraz-Mello\inst{1}
		\and Cristian Beaug\'e\inst{2}}
	\institute{Instituto de Astronomia, Geof\'isica e Ci\^encias Atmosf\'ericas, Universidade de S\~ao Paulo, Brasil \and
		Instituto de Astronom\'ia Te\'orica y Experimental, Observatorio Astron\'omico, Universidad Nacional de C\'ordoba, Argentina \\
		\email{sylvio@iag.usp.br}}
	\date{April 2025}

	\abstract
	{Observations with warm Spitzer and JWST revealed high and variable brightness in the planet 55 Cnc e.}
	{Inventory of the tidal effects on the rotational and orbital evolution of the planet 55 Cnc e enhanced by the nonzero orbital eccentricity.}
	{The creep-tide theory is used in simulations and dynamical analyses that explore the difficult trapping of the planet rotation in a 3:2 spin-orbit resonance and the most probable synchronization of the rotation.}
	{The strong tidal dissipation of energy, enhanced by the non-zero orbital eccentricity, may explain the observed brightness anomalies. However, the strong dissipation should also circularize the orbit. The observed non-zero eccentricity, if true, would indicate that an unknown planet in a close orbital resonance with 55 Cnc e perturbing the motion of this planet should exist. }	
	{}
	
	\titlerunning{Tidal heating of the exoplanet 55 Cnc e.}
	\authorrunning{S. Ferraz-Mello and C. Beaug\'e}
	\maketitle
	%
	
	\section{Introduction}
	
	This paper inventories the tidal effects on the rotational and orbital evolution of the planet 55 Cnc e. This planet has been the object of recent observations with warm Spitzer and JWST that revealed a high and variable brightness (Demory et al. 2016; Patel et al. 2024). This planet lies in an orbit very close to a solar-like host star and belongs to a non-compact system including four other planets. We discuss the dynamics of trapping the planet rotation in a 3:2 spin-orbit resonance and the more common alternative in which the rotation is synchronized with the orbital motion. We discuss the observed non-zero eccentricity of the system. 
	The fast circularization of any isolated ultra-short-period orbit in a few Myrs and the system's old age led us to think that the observed eccentricity, if true, is being forced by an unknown twin planet in a close orbital resonance with 55 Cnc e. 
	
	The creep-tide theory was chosen for this study because it does not introduce ad hoc models, has only one free parameter, and is valid for both gaseous and rocky bodies. The creep-tide theory arises from interpreting the matter displacements inside the body as a smooth radial flow without turbulence (low-Reynolds-number flow), namely, a Newtonian creep. The basic creep equation used to establish the theory is an approximate solution of the Navier-Stokes equation (Ferraz-Mello, 2013). However, the results shown in this paper are not exclusive to the creep-tide theory. They could also be obtained with the classical Darwinian theories, provided that appropriate ad hoc models were chosen to relate tidal lags and frequencies. 
	
	\section{Orbital evolution of a synchronized planet}\label{sec:sync}
	
\begin{figure}
		\centering
		\includegraphics[width=7cm]{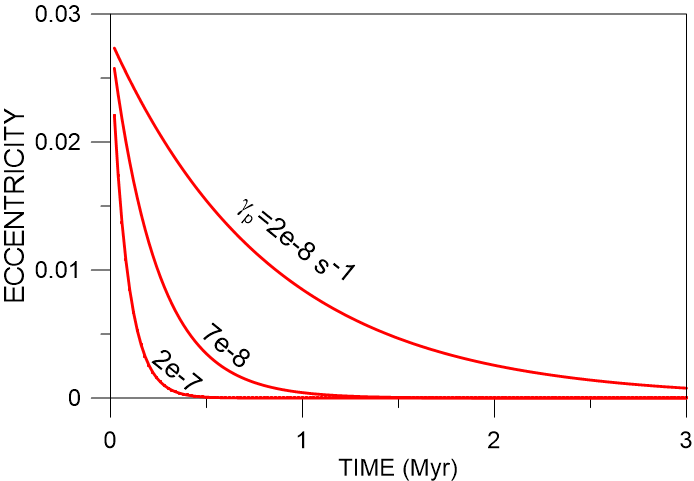}
		\caption{Evolution of the planet's orbital eccentricity in synchronous rotation. The labels indicate the planet's relaxation factor. $e_{(t=0)} = 0.028$. }
		\label{fig:v1Ecc}       
	\end{figure}

	We first study the system's evolution, assuming that the planetary rotation is synchronized. We consider that there are no other constraints and show that because of the tide on the planet, the orbital eccentricity is rapidly damped to zero (Fig. \ref{fig:v1Ecc}). We show the curves corresponding to three different values of the planet's relaxation factor, the largest of which, $\gamma = 2 \times 10^{-7}\ {\rm s}^{-1}$ is the value often adopted in the study of the solid Earth (see Ferraz-Mello, 2013). The other values shown correspond to harder bodies. We do not consider cases of a softer planet; in these cases, the damping is faster than the ones shown. 
	After the eccentricity is damped, the stellar tide dominates, and the planet falls toward the star. The fall on the star will occur in no more than 1 Gyr. The remaining lifetime depends on the star's relaxation factor. The adopted factors $\gamma_s = 50, 100, 200 {\rm s}^{-1}$ are based on the results of previous studies of the evolution of the rotation of host stars of the spectral types G7 -- K0 with large close-in companions (Ferraz-Mello et al. 2015).  
	
	\begin{figure}
		\centering
		\includegraphics[width=7cm]{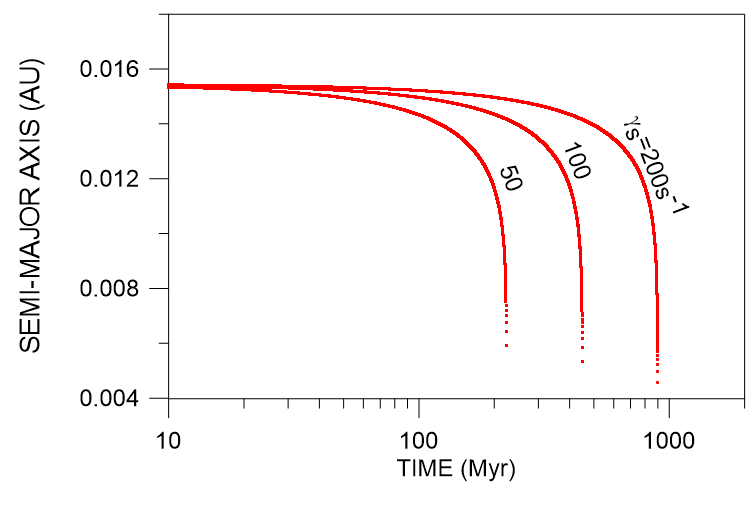}
		\caption{Evolution of the system with the planetary rotation synchronized with the orbital motion. The labels indicate the stellar tide relaxation factor. The planetary tide effect is negligible.  }
		\label{fig:v1Semi}       
	\end{figure}
	
	\begin{figure}
		\centering
		\includegraphics[width=7cm]{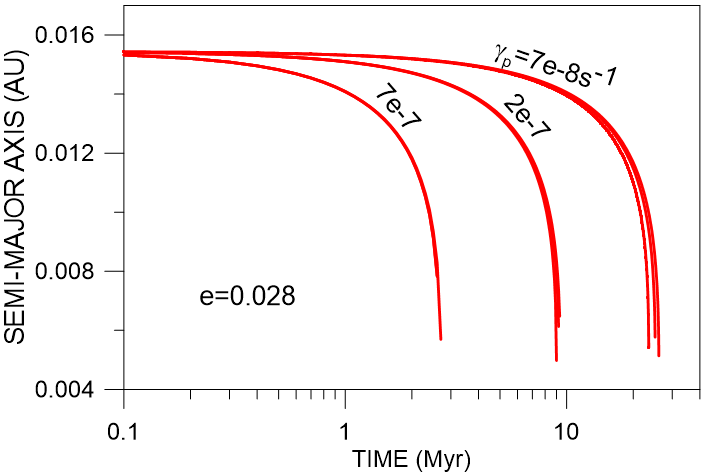}
		\caption{Evolution of the system with the planet rotation synchronized when the eccentricity is kept fixed at e=0.028. The labels show the planetary relaxation factor. Each line is a bundle of three solutions corresponding to the stellar relaxation factors 50, 100, and 200 s$^{-1}$. }
		\label{fig:v2Semi}       
	\end{figure}
	
	To complete this study, we show the evolution of the synchronous planet under the hypothesis that the observed eccentricity is forced by a third-body perturbation and so is not tidally damped (Fig. \ref{fig:v2Semi}). In this case, the tides on the planet are dominant and responsible for the fall of the planet in a much shorter time, less than 25 Myr. The stellar tide has only a small influence -- the curves shown are the superposition of the solutions corresponding to three different values of the stellar relaxation factor (the same appearing in Fig. \ref{fig:v1Semi}). The faster-falling case corresponds to a softer planet with $\gamma=7\times 10^{-7} {\rm s}^{-1}$ and was introduced to complete the range of values of $\gamma$ used for the solid Earth (see Ferraz-Mello, 2013). This solution would correspond to a decrease in the orbital period of $10^{-7} {\rm d}$ per year. This value is one order of magnitude smaller than the current precision of the period determinations.
	
	The interplay of the importance of the two tides in the above results may be understood by looking at the leading terms in the expressions of their contributions to the evolution of the semi-major axis. When the planet's rotation is assumed to be synchronous and the star is assumed to have a slow rotation, they are
	\begin{equation}
		\dot{a}_{\rm star} \simeq \frac
		{-6 k_s m_p R_s^5 n^2}
		{m_s a^4 \gamma_s},
	\end{equation}
	and
	\begin{equation}
		\dot{a}_{\rm planet} \simeq \frac
		{-75 k_p m_s R_p^5 e^2 \gamma_p}
		{4 m_p a^4},
	\end{equation}
	where the subscripts $s$ and $p$ indicate star and planet, respectively (Ferraz-Mello, 2013, 2022). The notations used are the usual ones: $m$ is mass, $R$ is the radius,  $a$ is the semimajor axis, $n$ is the mean motion, $e$ is eccentricity, and $k$ is the fluid Love number. 
	
	The ratio of the contributions of the two tides to the fall of the planet is given by
	\begin{equation}
		\frac{\dot{a}_p}{\dot{a}_s} =
		\mathbb{K} \frac{R_s}{R_p} \frac{\gamma_s \gamma_p}{n^2} e^2
		\label{eq:ratio}
	\end{equation}
	where $\mathbb{K}$ is a positive number of order ${\cal{O}}(1)$. The coefficient of $e^2$ in Eqn. \ref{eq:ratio} is much larger than 1 and compensates for the smallness of $e^2$, which makes the component due to the planetary tide dominate the variation of $a$. However, when $e \rightarrow 0$, this dominance ceases, and the evolution becomes dominated by the stellar tide (as seen in Fig. \ref{fig:v1Semi}). 
	
	\section{Orbital evolution of a planet initially in the 3:2 spin-orbit resonance}\label{sec:32}
	
	We repeat here the previous analysis but assume that the planet's rotation is initially trapped in the 3:2 resonance. \begin{figure}
		\centering
		\includegraphics[width=7cm]{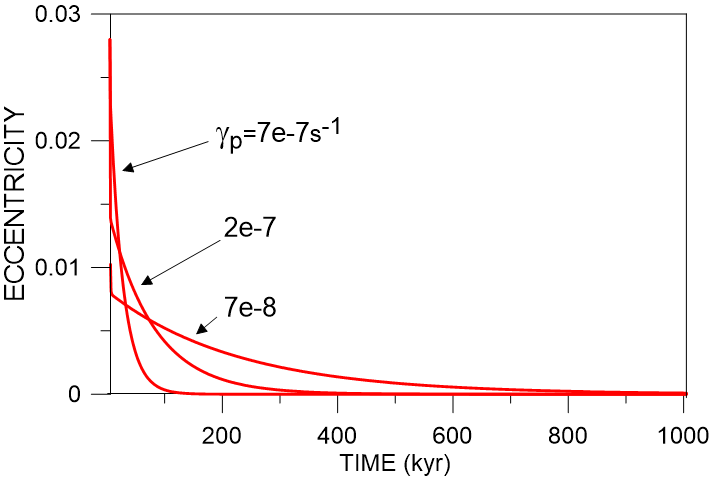}
		\caption{Evolution of the orbital eccentricity of the planet initially in 3:2 spin-orbit resonance. The labels indicate the planet's relaxation factor. }
		\label{fig:v4Ecc}       
	\end{figure}
	\begin{figure}
		\centering
		\includegraphics[width=7cm]{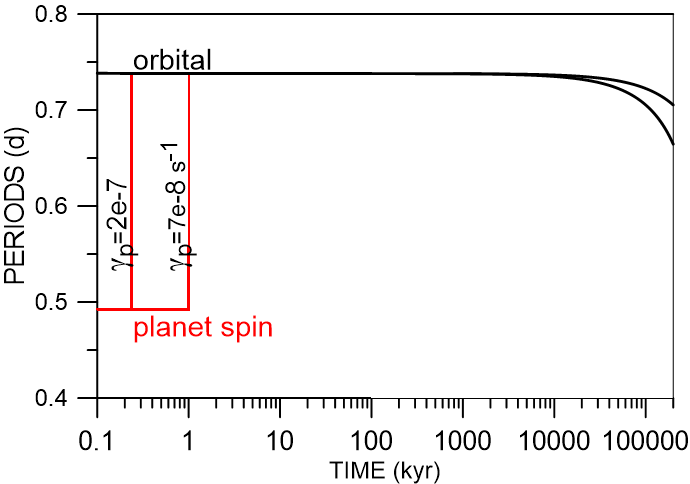}
		\caption{Evolution of the system with the planet rotation initially in 3:2 spin-orbit resonance for two different values of the stellar relaxation factor. The eccentricity is damped, and the rotation tends to be synchronous (the red and black curves merge).}
		\label{fig:v4PerNova}       
	\end{figure}
	
	Figures \ref{fig:v4Ecc} and \ref{fig:v4PerNova}\footnote{Figure \ref{fig:v4PerNova} replaces the one originally published, which was flawed due to a coding mistake. The text discussing the figure has been changed accordingly.} show rapid damping of the eccentricity and escape from the 3:2 spin-orbit resonance. We see two different phases in this process. First, the eccentricity is quickly damped to a small value. 
	When the decreasing eccentricity reaches the critical value corresponding to the spin-orbit resonance (see Appendix A.1), that stable stationary solution ceases to exist, and the rotation escapes the resonance and is driven to synchronization. 
	While trapped in the resonance, $\dot{a}>0$. 
	
	We recall that the contribution of the planetary tide to the variation of the semi-major axis when the planet's rotation is faster than the orbital motion is positive and, in the first approximation, independent of the eccentricity: 
	
	\begin{equation}
		\dot{a}_{\rm planet} \simeq \frac
		{+ 3 k_p m_s R_p^5 \gamma_p n \nu_p}
		{ m_p a^4 (\gamma_p^2+\nu_p^2)} \label{eq:dota}
	\end{equation}
	where $\nu_p=2(\Omega_p-n)$ is the so-called semi-diurnal frequency. This equation is not enough to understand the cases in the 3:2 spin-orbit resonance, but it applies to what is seen in Fig. \ref{fig:v4PerNova} because the planet escapes the resonance. 
	
	As before, to complete the study, we consider the evolution of the system when the planetary rotation is forced to remain in the 3:2 spin-orbit resonance and the observed eccentricity is forced by a third-body perturbation and not damped by the tides. 
	In this case, the planet eventually falls on the star in a time of a few hundred Myr. The evolution is dominated by the planetary tides but the simple approximation given by Eqn. \ref{eq:dota} is not enough to understand the results. This approximation fails in the neighborhood of the 3:2 spin-orbit resonance because of the smallness of $(\nu_p-n)$ in this neighborhood (see Appendix B).
	
	\begin{figure}
		\centering
		\includegraphics[width=7
		cm]{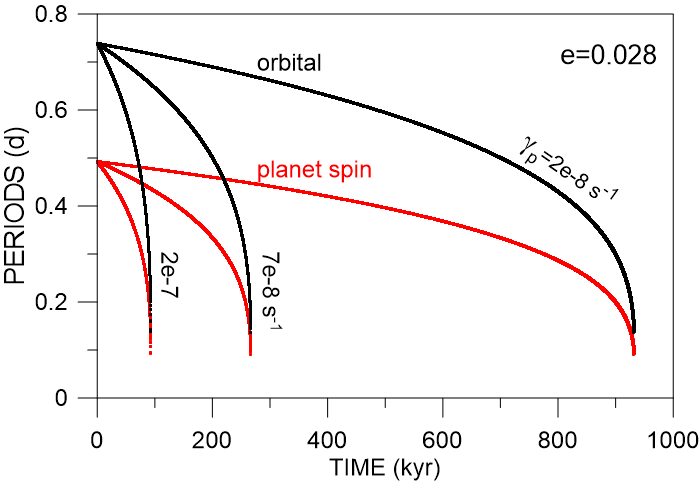}
		\caption{Evolution of the system with the planet rotation forced to remain in the 3:2 spin-orbit resonance and the orbital eccentricity kept fixed at e=0.028. The labels show the planetary relaxation factor. }
		\label{fig:v5Peri}       
	\end{figure}
	
	In these cases, the period decay is much faster than in the case of a synchronous planet. When the solution is forced to remain in the 3:2 spin-orbit resonance and $\gamma = 2\times 10^{-7}\ {\rm s}^{-1}$, we have a decrease in the orbital period of $ \sim 3 \times 10^{-6} {\rm d}$ per year. This value is of the same order of magnitude as the current precision of the period determinations and, if true, could be detected. We recall that $\dot{a}$ is proportional to the planetary relaxation factor, and thus a softer planet, with a larger relaxation factor, would be on the detection level.  
	
	\section{Heat Dissipation}
	
	The above results do not appear to favor the existence of a planet in a non-zero eccentricity orbit so close to the star. We either have a planet whose eccentricity is quickly damped to zero or a forced eccentricity accelerating the planet's fall on the star. In that case, orbital energy is being lost by the system. If we neglect the variations in the planet's geometry, the time derivatives in the orbital and rotational energies are given by:
	\begin{equation} 
		\dot{W}_{\rm orb}=\frac{\mathcal{G}m_s m_p}{2a^2}\dot{a}, \qquad
		\dot{W}_{rot pl}= I_p\Omega_p\dot{\Omega}_p
		\label{eq:diss}
	\end{equation}
	($I_p$ is the axial moment of inertia of the planet and $\Omega_p $ its rotation velocity).
	
	The variations may be determined from the quantities $\dot{P}$ obtained from the results shown in Fig. \ref{fig:v5Peri}. 
	As discussed in previous sections, the planetary tide is dominant when the eccentricity is forced to remain constant, and so the planet is the body responsible for the orbital energy loss. 
	
	In intermediate cases, corresponding to an Earth-like relaxation factor ($2\times 10^{-7} {\rm s}^{-1}$), we obtain, in the case of a synchronous rotation:
	$$\dot{W} \sim  -1.1 \times 10^{21}{\rm watts}. $$
	The mechanical energy lost is liberated within the planet and might flow through its surface at a rate $\sim 0.5  {\rm Mw/m}^2$. The amount of energy used to accelerate the planet's rotation is only a negligible fraction of it.
	
	This value is much larger than any other reported case. It is worth noting that if we use the formula for tidal heating based on the classical Darwin theory (see Barnes et al. 2009) and the quality factor $Q=500$, in the middle of the range accepted for the solid Earth, we obtain $ 0.45 {\rm Mw/m}^2$.  Quick et al. (2020) also reported a similar value.
	
	If the planet's rotation is trapped in the 3:2 resonance, this result is 100 times larger! 
	
	In general, the energy lost by the orbit may contribute to the heat release in both the star and the planet. In this case, calculations have shown that the variation $\dot{a}$ is due to the planetary tide.

	The above results favor a widespread volcanic activity as suggested by Winn et al.(2018) to explain the large irregular brightness in the dayside of the planet observed with Spitzer (Demory et al. 2016). If confirmed, it would explain the brightness anomalies by an intense volcanism of tidal origin and an incandescent planet.
	
	\section{Consequences of a lower eccentricity}
	
	There is not much we can do to get less dissipation on the planet. We need a slower orbital decay to reduce the dissipation. For example, we may assume a relaxation factor smaller than the chosen one. But this goes in the opposite direction to what we should expect. If the planet is hot, the fluidity may be greater, and consequently, the relaxation factor and the dissipation are expected to be rather larger! 
	
\begin{figure}
		\centering
		\includegraphics[width=7
		cm]{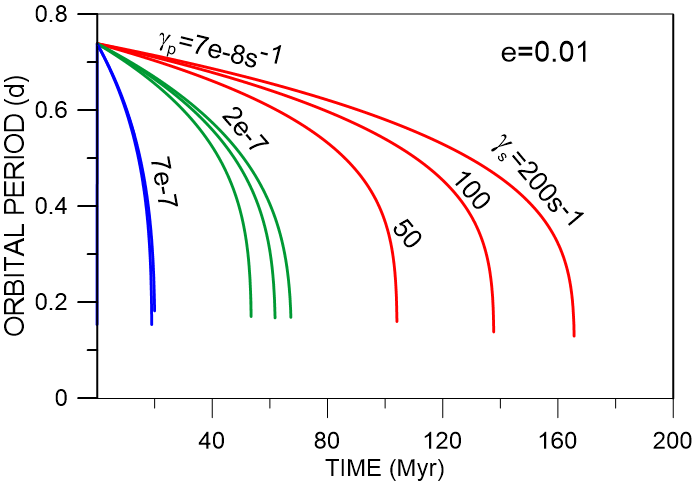}
		\caption{Evolution of the system with the planet rotation synchronized with the orbital motion, for fixed eccentricity at e=0.1 and relaxation factors as indicated by the labels. The same colors indicate solutions corresponding to the same planetary relaxation factor. Each bundle contains solutions corresponding to three different stellar relaxation factors. }
		\label{fig:v8Peri}       
	\end{figure}

	We may change the eccentricity. The nominal value has a large error bar: $0.028 \pm 0.018$. The minimum value in the $1\sigma$ confidence domain is $e=0.01$. With this lower eccentricity, no capture in the 3:2 spin-orbit resonance is possible, and synchronous rotation is the only possibility. Repeating what was done in Section \ref{sec:sync}, we obtain Fig. \ref{fig:v8Peri}.
	
	Since the eccentricity is smaller, the contribution of the planetary tide is not as dominant as when e=0.028. It still dominates the evolution, but the stellar tide also contributes. Then, for each planetary relaxation factor, we see 3 lines in the plot corresponding to the stellar relaxation factors 50, 100, and 200 s$^{-1}$.
	
	The decrease in the period when $\gamma_p=2\times 10^{-7}\ {\rm s}^{-1}$ is $3.5 \times 10^{-9} {\rm d}$ per year. The overall heat dissipation is now smaller than before but still large: 70 kw/m$^2$. 
	
	\section{The forced eccentricity}
	
	The last figure was done assuming that the observed eccentricity is forced and, therefore, cannot be damped by the tidal torques. This hypothesis arises from the fact that, otherwise, the orbit of 55 Cnc e would be rapidly circularized. We are maybe looking at this 10-Gyr-old system just when this final fast circularization occurs! However, it is more likely that this eccentricity results from perturbations by other bodies. The known planets cannot produce an eccentricity of 0.028 or even 0.01. The maximum evection-like perturbations found are thousands of times smaller. The same is true for the perturbations due to the star companion 55 Cnc B. Since tides do not allow an eccentricity to remain unaltered in a planet so close to the host star as 55 Cnc e, we are forced to assume that an unknown body is producing them. 
	
	We can, for instance, assume that it is located in the large gap between 55 Cnc e and its closest sibling, the hot Jupiter 55 Cnc b. The calculations can follow the same recipe used to study the forced eccentricity of LP 791-18 d (Ferraz-Mello et al. 2025). They show that a twin planet, with the same mass as 55 Cnc e, cannot produce a forced eccentricity in the confidence interval of the observed eccentricity unless it is in a close resonance with 55 Cnc e as the 2:1 or 3:2 period resonances. 
	Outside these resonances, it cannot produce the observed eccentricity.
	Unfortunately, there is no known way to detect such a planet. The TTVs due to such a companion would be too small to be observed: only 1 second.
	
	The other possibility is that this planet had a close approach to another body, in a higher orbit, and had its eccentricity excited by the encounter. If so, all simulations show that the orbit circularizes rapidly. This encounter should have occurred no more than 10 or 20 Myr ago. Although unexpected, these encounters are not rare. 
	
	Finally, the actual eccentricity may be closer to zero than the adopted value. The compatibility of circularization with the observations is in the $2\sigma$ range. 
	Since $\dot{a}_{p}$ is proportional to $e^2$, the planet's tidal dissipation would vanish, and the strong irradiation on the planet should explain the observed phenomena. 
	However, all other determinations (Baluev, 2015; Bourrier et al. 2018; Rosenthal et al. 2021) give values that are yet larger than the value e = 0.028 determined by Nelson et al. (2014).

	\section{Conclusion}
	
	This paper inventoried the tidal dynamics of planet 55 Cnc e, searching for facts that may help to understand the large brightness irregularities observed by warm Spitzer and JWST (Demory et al. 2016; Patel et al., 2024). The first point examined was the possibility of a capture of the rotation in a 3:2 spin-orbit resonance. With a nominal eccentricity of 0.028, this possibility exists. However, this is not a robust situation. The eccentricity is strongly affected by the planetary tides and decreases quickly, leading the planetary rotation to soon escape from the 3:2 resonance and be synchronized. After synchronization and circularization, the planetary tide ceases to dominate the system's evolution, and the weaker stellar tide becomes dominant. The planet may remain orbiting the star for some hundred million years, the exact remaining time until the fall being dependent on the stellar relaxation factor (Fig. \ref{fig:v1Semi}).  
	
	The only way to avoid rapid circularization of the orbit is the eccentricity being forced by a third body. However, the other planets in the system are far away and cannot force an eccentricity larger than a few thousandths. If so, this third body must be an unknown one. But almost all initial conditions between 55 Cnc e and its closest sibling lead to rapid circularization. 
	The only exceptions occur if the third body is assumed to be very close to the 2:1 or 3:2 period resonances with 55 Cnc e. 
	
	The presence of a forced eccentricity means that circularization of the orbit did not happen. However, the planet is too close to the star, and the forced eccentricity contributes to increasing the effects of the planetary tide and accelerates the star's rate of fall toward the star.
	
	The orbital energy decreases quickly and may be dissipated inside the planet. If the dissipated energy flows through the planet's surface, it corresponds to many kilowatts per square meter; the actual value depends on the forced eccentricity and the planet's relaxation factor. A so high heat flow has not been found in any planet studied before, but it must be normal for planets with non-zero eccentricity so close to a solar-like host star. The volcanism it may induce can be responsible for the irregular brightness observed by warm Spitzer (Winn et al., 2016). 
	
	All calculations show that this planet is falling onto the star in, at most, some hundreds of millions of years. 
	
	The calculations were done using the creep-tide theory equations expanded up to the 6th power of eccentricities. The simplified formulas included in the text were only used to identify the individual roles of the stellar and planetary tides in each case. 
	
	\begin{acknowledgements}
		This investigation was sponsored by CNPq (Proc. 303540/2020-6) and FAPESP (Procs. 2016/13750-6 ref. PLATO mission).  
	\end{acknowledgements}

	\begin{appendix}

		\section{Spin-orbit resonance trapping dynamics}\label{sec:trap}
		
		Usual studies of spin-orbit resonance consider an asymmetric companion (see Celletti, 2010; Callegari and Rodr\'{\i}guez, 2013; Pinzari et al., 2024). The equatorial asymmetry is responsible for a force that counterbalances the tidal torques, allowing the rotation of the body to be synchronized with the orbital motion (Ferraz-Mello et al. 2008; Rodr\'{\i}guez et al. 2012). The differential equations are pendulum-like, second-order differential equations. However, if an asymmetry does not exist a priori, the rotational dynamics is rather ruled by a first-order differential equation (Correia et al, 2014; Ferraz-Mello, 2015) of the form
		\begin{equation}\label{eq:Omegadot}
			\langle \dot\Omega \rangle = -\mathcal{A} \sum_{k \in \mathbb{Z}} E_{2,k}^2\frac{\gamma(\nu+kn)}{\gamma^2+(\nu+kn)^2}
		\end{equation}
		(Ferraz-Mello, 2013; Ferraz-Mello et al. 2022)
		where $\mathcal{A}$ is a positive coefficient depending on the dynamical parameters of the system, $\gamma$ is the relaxation factor of the body, $n$ the orbital mean motion,  and $E_{2,k}$ known functions of the orbital eccentricity (Cayley coefficients).
		
		\begin{figure}[h]
			\centering
			\includegraphics[width=7cm]{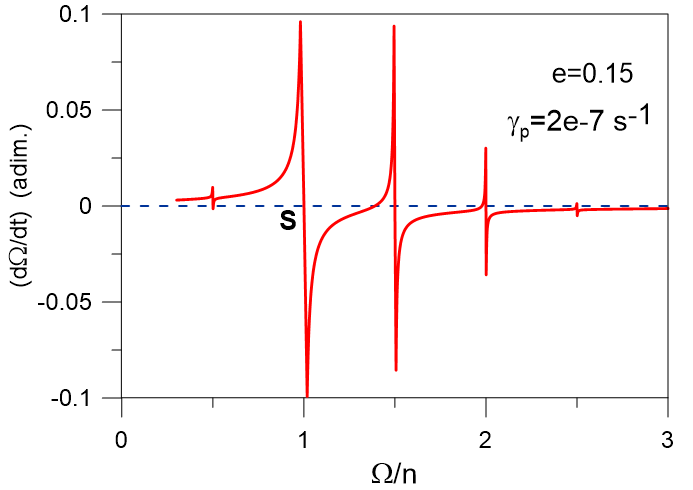}
			\caption{Phase plane  $(\Omega, \langle \dot\Omega \rangle)$ of Eqn. \ref{eq:Omegadot} in the case of a hypothetical planet with the same characteristics as 55 Cnc e and an enhanced orbital eccentricity $e=0.15$. The point $\mathbb{S}$ is the locus of the synchronous solution.}
			\label{fig:Omega1}       
		\end{figure}
		
		Figure \ref{fig:Omega1} shows the phase plane of Eqn. \ref{eq:Omegadot} in one special case chosen for its rich features. In this case, the parameters are those of the system formed by 55 Cnc e and its host star with a planetary relaxation factor $\gamma = 2\times 10^{-7}\ {\rm s}^{-1}$ -- a relaxation factor of the order of the solid Earth's relaxation factor (see Ferraz-Mello, 2013). The eccentricity was arbitrarily fixed at e = 0.15, a value unrealistic but good to emphasize the main features of the equation.
		
		\begin{figure}
			\centering
			\includegraphics[width=3cm]{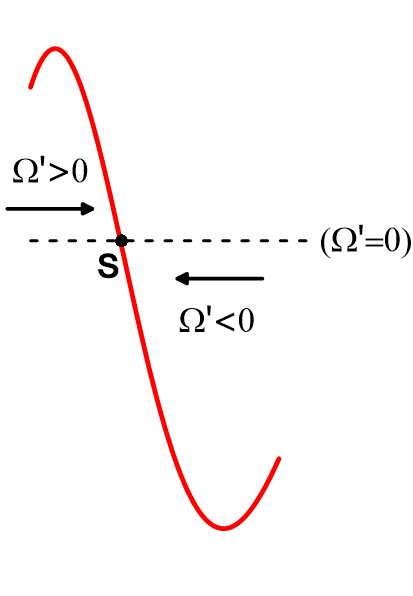}
			\caption{Stable stationary solution ($\mathbb{S}$).}
			\label{fig:plot1}       
		\end{figure}
		
		The points where the curve crosses the axis $\langle \dot\Omega \rangle = 0$ are equilibrium points of the system, a.k.a. stationary solutions of Eqn. \ref{eq:Omegadot}. When the curve crosses the axis in the descending direction (the derivative of $\langle \dot\Omega \rangle$ is negative), the solution is stable. The velocity is positive on the left of the crossing point $\mathbb{S}$ and negative on its right (see Fig. \ref{fig:plot1}). In contrast, the corresponding stationary solution is unstable if the curve is ascending at the crossing point (the derivative of $\langle \dot\Omega \rangle$ is positive).  
		
		In Fig. \ref{fig:Omega1} one may see 5 stable stationary solutions, at the frequencies $\Omega=\frac{kn}{2}$ $(k=1,2,3,4,5)$. The two corresponding to $k=1$ and $k=5$ would require a zoom to be seen, but the other three are evident. In its evolution, the planet's rotation may be trapped in any of these solutions. If the planet starts as a fast rotator, it will evolve from the right to the left and will be trapped in the 5/2 orbital resonance. If, because of a variation in its eccentricity, it succeeds in escaping to this resonance, it will be trapped in the next one. The pseudo-synchronous stationary solution at $\mathbb{S}$ is the only one for which there is no escape route.
		We say pseudo-synchronous because the point $\mathbb{S}$ is not exactly at $\Omega=n$, but slightly at its right (see Ferraz-Mello et al. 2008; Ferraz-Mello, 2013).
		
		\subsection{The trapping in the 3:2 spin-orbit resonance.}
		
		\begin{figure}
				\centering
			\includegraphics[width=7cm]{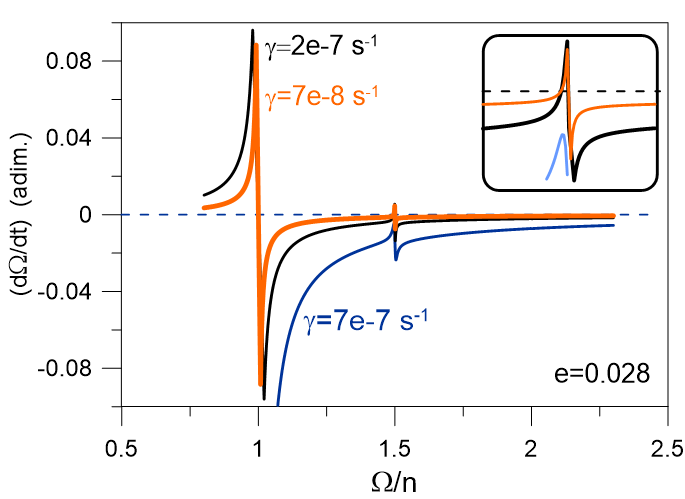}
			\caption{Phase plane  $(\Omega, \langle \dot\Omega \rangle)$ of Eqn. \ref{eq:Omegadot} in the case of the real 55 Cnc e, with three different values of the relaxation factor. The inset box shows a zoom of the neighborhood of the stable equilibrium corresponding to the 3:2 spin-orbit resonance.}
			\label{fig:zoom2}       
		\end{figure}
		
		Let us now consider the phase plane $(\Omega, \langle \dot\Omega \rangle)$ of Eqn. \ref{eq:Omegadot} in some cases akin to the actual 55 Cnc e. Fig. \ref{fig:zoom2} shows the curve corresponding to the observed eccentricity of 55 Cnc e and three different values for the relaxation factor: $\gamma = 7\times 10^{-8}\ {\rm s}^{-1}$, $\gamma = 2\times 10^{-7}\ {\rm s}^{-1}$, and $\gamma = 7 \times 10^{-7}\ {\rm s}^{-1}$. These values are of the same order as the relaxation factor of the solid Earth. The inset on the top of the figure shows that in the softer case ($\gamma = 7\times 10^{-7}\ {\rm s}^{-1}$), the crest at the 3:2 resonance does not cross the dashed line, which means that, in this case, the corresponding stationary solution does not exist. The planet may be trapped in this resonance in the harder cases, but not in the softer ones. 
		
		No crest appears at $\Omega=2n$, showing that no other non-synchronous stationary solution exists beyond 3:2 at the current eccentricity of the planet.
		
\begin{figure}
				\centering
			\includegraphics[width=7cm]{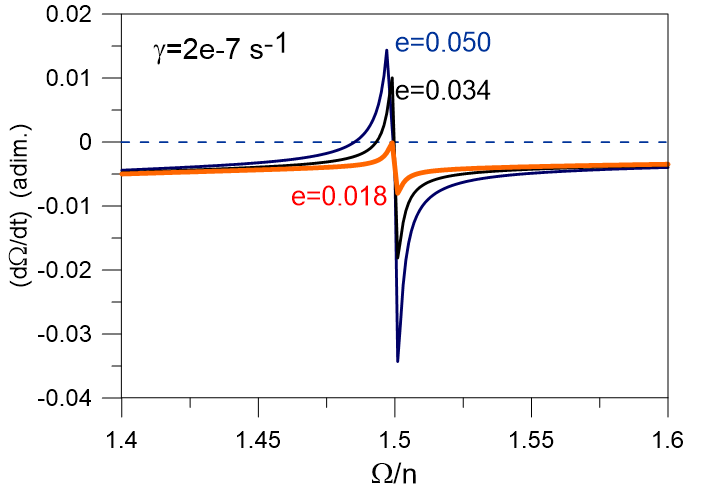}
			\caption{Phase diagram  $(\Omega, \langle \dot\Omega \rangle)$ of Eqn. \ref{eq:Omegadot} for three different values of the orbital eccentricity and the relaxation factor 
				$\gamma = 2\times 10^{-7}\ {\rm s}^{-1}$. The labels indicate the used eccentricities.}
			\label{fig:zoom3}       
		\end{figure}

\begin{figure}
			\centering
			\includegraphics[width=7cm]{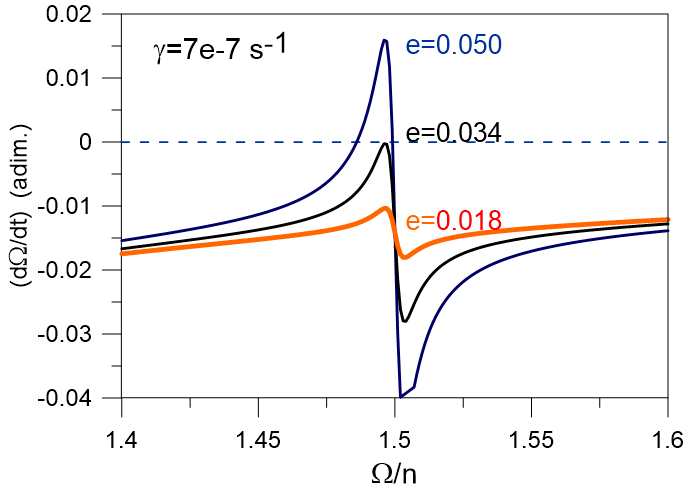}
			\caption{Same as Fig. \ref{fig:zoom3} for the case of a larger relaxation factor: $\gamma = 7\times 10^{-7}\ {\rm s}^{-1}$ (a softer planet).}
			\label{fig:zoom4}       
		\end{figure}

		In the other two figures (Figs. \ref{fig:zoom3} and \ref{fig:zoom4}), we discuss the dependence of the trapping on the orbital eccentricities to know the limiting values necessary to have the spin-orbit 3:2 resonance. 
		
		Fig. \ref{fig:zoom3} shows the phase diagram for the relaxation factor 
		$\gamma = 2\times 10^{-7}\ {\rm s}^{-1}$ and the eccentricities 0.018, 0.034, and 0.050. We see that in all cases the curves intersect the axis $\dot{\Omega} = 0$. When e=0.018, we would need a zoom to see if it actually crosses the axis or not. But this detail is not important. What matters is that close to e=0.018 we have the limit necessary for trapping in the stationary 3:2 solution. If, in some instant, because of any perturbations, the eccentricity falls below that limit, the solution will escape the resonance and evolve toward synchronization. And there is no way to come back!

		Fig. \ref{fig:zoom4} is the same as above for a softer planet\footnote{We remind that the relaxation factor $\gamma$ is inversely proportional to the dynamical viscosity (Darwin, 1879; Ferraz-Mello, 2013).} with $\gamma = 7\times 10^{-7}\ {\rm s}^{-1}$ and the same eccentricities as before. In this case, the lower trapping limit is $e \sim 0.034$. With this relaxation factor, the planet's rotation cannot be trapped in the 3:2 spin-orbit resonance, and the system should have evolved to the pseudo-synchronous solution. Conversely, if we prove that the rotation is trapped in the 3:2 resonance, we may say that the relaxation factor is smaller than $7\times 10^{-7}\ {\rm s}^{-1}$.
		
		It is noteworthy that the limits found in Figs. \ref{fig:zoom3} and \ref{fig:zoom4} agree perfectly with the predictions given by the condition
		\begin{equation}
			e^2 \ge \frac{8\gamma}{49 n} \approx \frac{\gamma}{6n}
			\qquad\qquad (\gamma \ll n)
		\end{equation}
		found by Correia et al. (2014) for the minimum eccentricity necessary for the existence of the stable equilibrium at the 3:2 spin-orbit resonance.
		
		\section{Variation of the semi-major axis}
		
		Equation \ref{eq:dota} is common to all tidal theories and generally works fine. It may be quickly interpreted when we note that $\dot{a}_p$ has the same sign as $\nu_p$. So $\dot{a}$ is positive when $\Omega_p>n$, and negative otherwise. The same is true with $\dot{a}_s$. So, as the Galilean satellites orbit a fast-rotating planet, they recede from it, whereas close-in planets generally orbit slow-rotating stars and thus fall on the star. However, this classical statement is only true if the considered satellites or planets are in almost circular orbits. 
		
		Equation \ref{eq:dota} is the limit of a more general equation involving eccentricities when $e \rightarrow 0$. However, in the vicinity of the 3:2 spin-orbit resonance, this limit is singular and needs special treatment.
		
		The full expression, in second order, is
		\begin{eqnarray}
			\dot{a}_{\rm planet} & \simeq & \frac
			{+ 3 k_p m_s R_p^5 n }
			{ m_p a^4}
			\left( (1-5e^2) \Psi(\nu_p)- \phantom{\frac{e^2}{2}}\right. \\
			&& \qquad  \left.  \frac{3e^2}{4} \Psi(n) + \frac{e^2}{8}\Psi(\nu_p+n)+\frac{147e^2}{8}\Psi(\nu_p-n) \right) \nonumber \label{eq:IAU}
		\end{eqnarray}
		(Ferraz-Mello, 2022) where $\Psi(x)$ is the Maxwell operation
		\begin{equation}
			\Psi(X)=\left(\frac{\gamma_p}{X}+\frac{X}{\gamma_p}\right)^{-1}.
		\end{equation}
		The first three terms in brackets do not present difficulties. In these cases, we have $\gamma_p \ll |X|$ and may write $\Psi(X)\simeq \frac{\gamma_p}{X}$. However, the fourth term needs a more detailed calculation. 
		If the eccentricity is large enough to allow the crest formed near the point $\Omega_p=\frac{3n}{2}$ (see Fig. \ref{fig:zoom4}) to cross the axis $\dot{\Omega}_p=0$, the equation $\dot{\Omega}_p=0$ has two real roots. 
		They are $\nu_p-n \simeq -\frac{49}{4}ne^2$ (unstable) and $\nu_p-n \simeq -\frac{4\gamma_p^2}{49ne^2}$ (stable). If $\gamma_p \ll \frac{49}{4}ne^2$, we get the approximate result $\frac{147e^2}{8}\Psi(\nu_p-n) \simeq -\frac{3}{2} $, which makes $\dot{a}_p<0$. Clearly, this result is not true for a very small eccentricity, but it is valid in the case under study and explains the behavior observed by the evolution curves in Fig. \ref{fig:v5Peri}. 
		It is not visible in Fig. \ref{fig:v4PerNova}, notwithstanding the fact that the initial rotation is in the 3:2 spin-orbit resonance, because the escape of this condition is very fast. 
		
		\section{Used parameters} 
		When not otherwise stated, the initial conditions in all simulations correspond to the nominal orbit of the planet. 
		The masses, radii, and orbital elements used are those given in the catalog of exoplanetary systems at exoplanets.eu.  The other parameters used are given in table \ref{tab1}.
		
		\begin{table}[h]
			
			\centering
			\caption{Stellar and planetary parameters adopted}
			\label{tab1}
			\begin{tabular}{l@{\qquad}cc}
				\hline
				& Star  &  Planet  \\
				\hline
				Mass & 0.9 M$_\odot$ & 0.02703 M$_{\rm J}$ \\
				Radius & 0.963 R$_\odot$ & 0.1737 R$_{\rm J}$\\
				Relaxation factor (s$^{-1}$)   & $50-200$  \qquad &  $0.2 - 7 \times 10^{-7}$   \\
				Moment of Inertia ($\times\ mR^2$) \qquad   & 0.07   & 0.33 \\
				Fluid Love number    & 0.26   & 1.28   \\
				Rotation Period (days)       &  42.7  &  $-$\\
				Semi-major axis (AU) && 0.015439 \\
				Eccentricity && 0.028 \\
				\hline
			\end{tabular}
		\end{table}

	\end{appendix}
	
	\end{document}